\documentclass[useAMS,usenatbib]{mn2e}
\usepackage{epsfig}
\usepackage{amsmath, amssymb,bm}

\title[Alignment and Precession of Black Holes]{Alignment and Precession of a Black Hole with a Warped Accretion Disc}

\author[R. G. Martin, J. E. Pringle and C. A. Tout]{Rebecca G. Martin,
  J. E. Pringle and Christopher A. Tout \\ University of Cambridge,
  Institute of Astronomy, The Observatories, Madingley Road, Cambridge
  CB3 0HA\\}
\begin{document}

\date{}

\pagerange{\pageref{firstpage}--\pageref{lastpage}} 
\pubyear{2007}
\maketitle

\label{firstpage}

\begin{abstract}
  We consider the shape of an accretion disc whose outer regions are
  misaligned with the spin axis of a central black hole and calculate
  the steady state form of the warped disc in the case where the
  viscosity and surface densities are power laws in the distance from
  the central black hole. We discuss the shape of the resulting disc
  in both the frame of the black hole and that of the outer disc. We
  note that some parts of the disc and also any companion star maybe
  shadowed from the central regions by the warp. We compute the torque
  on the black hole caused by the Lense-Thirring precession and hence
  compute the alignment and precession timescales. We generalise the
  case with viscosity and hence surface density independent of radius
  to more realistic density distributions for which the surface
  density is a decreasing function of radius. We find that the
  alignment timescale does not change greatly but the precession
  timescale is more sensitive. We also determine the effect on this
  timescale if we truncate the disc. For a given truncation radius,
  the the timescales are less affected for more sharply falling
  density distributions.
\end{abstract}


\section{Introduction}

Observations indicate that accretion discs around black holes can be
warped. Warped discs have been observed in active galactic nuclei
(AGN) by water maser observations in NGC 4258 \citep{H96} and in the
Circinus galaxy \citep{G03}. A warped inner accretion disc might
explain why radio jets from AGN are not perpendicular to the plane of
the Galactic disc \citep{K00, S02}.

The two X-ray binaries GRO~J~1655-40 and SAX~J1819-2525 have also been
observed to have jets misaligned with their orbital planes. For
example, GRO~J~1655-40 appears to have a binary orbit at $70^\circ$
\citep{G01} and jet inclination at $85^\circ$ \citep{H95}.  This
implies that there is a misalignment of at least $15^\circ $ between
the inclination of the black hole and the outer parts of the accretion
disc.

We consider a system with an accretion disc around a spinning black
hole. The black hole spin is misaligned with the outer parts of the
disc which we assume to be fixed by the plane of a binary companion.
Lense-Thirring precession drives a warp in the disc which reaches a
steady state when the inner parts are aligned with the black hole by
the \cite{BP} effect.

\cite{SF} calculated the shape of the steady disc and the timescale
for the black hole to align on the assumption that the warping is
gradual and that the viscosities and the surface density of the disc
are independent of radius. Because in their analysis the orientation
of the outer disc is fixed, the torque between the disc and the hole
makes the black hole precess and makes its spin align with that of the
disc \citep{King05}. \cite{SF} find that the alignment timescale and
precession timescale are about the same in this case and this has been
illustrated with numerical simulations by \cite{LP06}. However, in
more realistic discs the surface density is a decreasing function of
radius. For idealised discs in which in which shear viscosity $\nu_1$
varies as a power law, $\nu_1 \propto R^\beta$, in radius $R$ the
steady state surface density $\Sigma$ obeys $\Sigma \propto
R^{-\beta}$ because $\nu_1 \Sigma$ tends to a constant far from the
inner edge \citep{P81}.  Typically the power $\beta$ lies in the range
$0\le \beta \le 2$. 

\cite{NA99} discuss how the alignment timescale depends on $\beta$ and
suggest on dimensional grounds that the timescale increases by about a
factor of $10$ as $\beta$ changes from 0 to 1.5. We compute the shape
of the disc but our analysis differs from that of \cite{NA99} in that
we compute how the precesssion and alignment timescales vary with the
density distribution ($\beta$) when the warp radius and the accretion
rate are fixed. We find that the alignment timescale does not vary
strongly with $\beta$ but that the precession timescale is more
sensitive..

\section{Steady State Solution}

Our analysis follows that of \cite{SF}.  We consider the disc to be
made up of annuli of width $dR$ and mass $2\pi \Sigma R dR$ at radius
$R$ from the central star of mass $M$ with surface density
$\Sigma(R,t)$ at time $t$ and with angular momentum
$\bm{L}=(GMR)^{1/2}\Sigma \bm{l}=L\bm{l}$.  The unit vector describing
the direction of the angular momentum of a disc annulus is given by
$\bm{l}=(l_x,l_y,l_z)$ with $|\bm{l}|=1$.

Like \cite{SF} we derive a solution for the steady state disc profile
described by $W=l_{x}+il_{y}$.  We use equation (2.8) of \cite{P92}
setting $\partial \bm{L}/ \partial t = 0$ and adding a term to
describe the Lense-Thirring precession to give
\begin{align}
0=&\frac{1}{R}\frac{\partial}{\partial R}\left[ \left( \frac{3R}{L} \frac{\partial}{\partial R}(\nu_1 L)
  -\frac{3}{2}\nu_1\right)\bm{L}+\frac{1}{2}\nu_2RL\frac{\partial \bm{l}}{\partial R}\right] \cr
 & + \frac{\bm{\omega_{\rm p}} \times \bm{L}}{R^3}.
\label{maineq}
\end{align}
There are two viscosities, $\nu_1$ corresponds to the azimuthal shear
(the viscosity normally associated with accretion discs) and $\nu_2$
corresponds to the vertical shear in the disc which smoothes out the
twist. The second viscosity acts when the disc is non-planar. The
Lense-Thirring precession is given by
\begin{equation}
\bm{\omega_{\rm p}} =\frac{2G\bm{J}}{c^2},
\label{omegap}
\end{equation}
\citep{KP85} where the angular momentum of the black hole
$\bm{J}=J\bm{j}$ with $\bm{j}=(j_x,j_y,j_z)$ and $|\bm{j}|=1$ can be
expressed in terms of the dimensionless spin parameter $a$ such that
\begin{equation}
J=acM\left(\frac{GM}{c^2}\right).
\label{angmom}
\end{equation}

We take both viscosities to have power law form so that
\begin{equation}
\nu_1=\nu_{10}\left(\frac{R}{R_0}\right)^\beta ~~~ {\rm and} ~~~ 
\nu_2=\nu_{20}\left(\frac{R}{R_0}\right)^\gamma ,
\end{equation}
where $\nu_{10}$, $\nu_{20}$, $\beta$ and $\gamma$ are all constants
and $R_0$ is some fixed radius. The surface density is
\begin{equation}
\Sigma= \Sigma_0 \left(\frac{R}{R_0}\right)^{-\beta}.
\end{equation}
We take the scalar product of equation~(\ref{maineq}) with $\bm{l}$
and find
\begin{equation}
0=\frac{1}{R}\frac{\partial}{\partial R}\left[3R\frac{\partial}{\partial R}(\nu_1 L)-\frac{3}{2}\nu_1L\right]
\label{int}
\end{equation}
because $\bm{l}.\partial \bm{l}/\partial R=0$ when $|\bm{l}|=1$. We
assume the warp is gradual enough that we can neglect the non-linear
term $\bm{l}.\partial^2\bm{l}/\partial R^2= -\left|\partial \bm{l} /
  \partial R \right|^2$. We consider the effects of neglecting this
term at the end of this section. This equation has the solution
\begin{equation}
\nu_1 L= C_0R^{1/2}+C_1
\end{equation}
where $C_0$ and $C_1$ are constants.  We set $L=0$ at $R=0$ and
$L=(GMR)^{\frac{1}{2}}\Sigma$ so $C_1=0$ and
\begin{equation}
L=(GMR)^{\frac{1}{2}}\Sigma_0\left(\frac{R}{R_0}\right)^{-\beta}
\label{modL}
\end{equation}
in the steady state as in the flat case.

Substituting equation~(\ref{modL}) into equation~(\ref{maineq}) we
find
\begin{equation}
-\frac{\bm{\omega_{\rm p}}\times \bm{l}}{R^2} L 
= \frac{\partial}{\partial R}\left[ \frac{1}{2}R\nu_2 L \frac{\partial \bm{l}}{\partial R}\right].
\label{balance}
\end{equation}
We work in the frame of the black hole where $\bm{J}/J=(0,0,1)$ so
that $\bm{\omega_{\rm p}}=(0,0,\omega_{\rm p})$ and $\bm{\omega_{\rm
    p}}\times \bm{l} = (-\omega_{\rm p}l_y,\omega_{\rm p}l_x,0)$.

We add the $x$-component of equation~(\ref{balance}) to $i$ times the
$y$-component, where $i = \sqrt{-1}$  and set
$W=l_x+il_y$ to obtain
\begin{equation}
\kappa R^{-\frac{3}{2}-\beta}W=\frac{d}{dR}\left[R^{\gamma+\frac{3}{2}-\beta}\frac{dW}{dR}\right],
\end{equation}
where
\begin{equation}
\kappa =-\frac{2 i \omega_{\rm p}}{\nu_{20}}R_{0}^{\gamma},
\label{kappa}
\end{equation}
so that
\begin{equation}
\kappa ^{\frac{1}{2}}=\pm(1-i)\left(\frac{\omega_{\rm p}}{\nu_{20}}\right)^{\frac{1}{2}}R_0^{\frac{\gamma}{2}}.
\label{kappahalf}
\end{equation}

To simplify the analysis we let $\beta=\gamma$ so that the two
viscosities obey the same power law and thus the ratio $\nu_1/\nu_2$
is independent of radius (c.f. Lodato \& Pringle, 2007).  Setting
$x=R^{-\frac{1}{2}(1+\beta)}$ we find
\begin{align}
\kappa x^{\frac{3+2\beta}{1+\beta}}W=\frac{(1+\beta)^2}{4}x^{\frac{3+\beta}{1+\beta}} \frac{d}{dx}\left( x^{\frac{\beta}{1+\beta}}\frac{dW}{dx}\right)
\end{align}
and setting $W(R)=R^{-\frac{1}{4}}V(R)=x^{\frac{1}{ 2(1+\beta)}}V(x)$
we obtain
\begin{align}
x^2\frac{d^2V}{dx^2}&+x\frac{dV}{dx}\cr
&-\left(\frac{1}{(2(1+\beta))^2}+\left(\frac{2\kappa^{\frac{1}{2}}}{1+\beta}\right)^2x^2\right)V=0.
\end{align}
This is a modified Bessel equation with solution
\begin{equation}
V=AI_{\frac{1}{2(1+\beta)}}\left(\frac{2}{1+\beta}\kappa^{\frac{1}{2}}x\right) +BK_{\frac{1}{2(1+\beta)}}\left(\frac{2}{1+\beta}\kappa^{\frac{1}{2}}x\right),
\end{equation}
where $I_\nu(z)$ and $K_\nu(z)$ are the modified Bessel functions of
the first and second kind respectively and $A$ and $B$ are constants
to be determined.  

We know that $W\rightarrow 0$ as $R\rightarrow 0$ because the inner
disc is aligned with the spin of the black hole. Thus we take $\kappa$
to have a positive real part and $A=0$ so that the full solution is
\begin{equation}
W=B\left(\frac{R}{R_0}\right)^{-\frac{1}{4}}K_{\frac{1}{2(1+\beta)}}\left(\frac{2}{1+\beta}\kappa^{\frac{1}{2}}R^{-\frac{1}{2}(1+\beta)}\right).
\label{W1}
\end{equation}
In order to find $B$ we need to consider what happens as $R\rightarrow
\infty$.  At large radius the disc tilt is taken to be fixed and thus
we let $W\rightarrow W_\infty$ which is a constant. Thus as $R
\rightarrow \infty$ we find
\begin{equation}
W\rightarrow \frac{BR_0^{\frac{1}{4}}}{2}\Gamma \left(\frac{1}{2(1+\beta)} \right)  \left(\frac{\kappa^{\frac{1}{2}}}{1+\beta}\right)^{-\frac{1}{2(1+\beta)}} =W_\infty,
\end{equation}
where we have made use of the relation
\begin{equation}
K_{\nu}(x)\sim \frac{\Gamma(\nu)}{2}\left(\frac{x}{2}\right)^{-\nu},
\label{Ksmallx}
\end{equation}
as $x \rightarrow 0$.
Rearranging we now find
\begin{equation}
B=\frac{2}{\Gamma\left(\frac{1}{2(1+\beta)}\right)} R_0^{-\frac{1}{4}}\left(\frac{\kappa^{\frac{1}{2}}}{1+\beta}\right)^{\frac{1}{2(1+\beta)}}W_\infty
\label{B}
\end{equation}
and
\begin{align}
W=& W_\infty\frac{2}{\Gamma \left(\frac{1}{2(1+\beta)}\right)}\left(\frac{\kappa^{\frac{1}{2}}}{1+\beta}\right)^{\frac{1}{2(1+\beta)}} \cr
&\times R^{-\frac{1}{4}}K_{\frac{1}{2(1+\beta)}}\left(\frac{2}{1+\beta}\kappa^{\frac{1}{2}}R^{-\frac{1}{2}(1+\beta)}\right).
\label{W2}
\end{align}
We note that if $\beta=0$  this reduces to
\begin{equation}
W=W_\infty \exp \left[-2(1-i)\left(\frac{\omega_{\rm
      p}}{\nu_{20}R}\right)^{1/2}\right],
\end{equation}
where we have made use of the identities
$K_{1/2}(z)=e^{-z}\sqrt{\pi/(2z)}$ and $\Gamma (1/2)=\sqrt{\pi}$.
This is the solution found by \cite{SF}.

The radius where the warp in the disc typically occurs, $R_{\rm warp}$
\citep{SF}, can be found by balancing the terms on either side of
equation~(\ref{balance}). We find
\begin{equation}
R_{\rm warp} = \left(\frac{2\omega_{\rm p}}{\nu_{20}}R_0^\beta
\right)^{1/(1+\beta)}.
\end{equation}
In order to compare different power laws for viscosity in a reasonable
way we want to keep $\nu_1$, $\nu_2$ and $\Sigma$ the same at $R_{\rm
  warp}$ where the torques are greatest. We therefore set
\begin{equation}
R_0=R_{\rm warp}=\frac{2\omega_{\rm p}}{\nu_{20}}  =\frac{4aG^2M^2}{\nu_{20}c^3}.
\end{equation}
and note that $\nu_{20}$ now corresponds to the value of $\nu_2$ at
the radius where the disc is warped.  Equation~(\ref{W2}) with
$\kappa=-i R_{\rm warp}^{1+\beta}$ becomes
\begin{align}
W=&\frac{2W_\infty}{\Gamma \left(\frac{1}{2(1+\beta)}\right)}\frac{(-i)^{\frac{1}{4(1+\beta)}}}{ (1+\beta)^{\frac{1}{2(1+\beta)}}}\left(\frac{R_{\rm warp}}{R}\right)^{1/4} \cr
&\times K_{\frac{1}{2(1+\beta)}}\left(\frac{\sqrt{2}}{1+\beta}(1-i)\left(\frac{R}{R_{\rm warp}}\right)^{-\frac{1+\beta}{2}}\right).
\label{Wwarp}
\end{align}
Note that we choose the negative root of $-i$ in
equation~(\ref{kappahalf}) because we want the real part of the
argument of the Bessel function to be positive. 
We note that in this solution we have a term of the form
\begin{equation}
K_\nu (e^{-i\pi/4}x)=e^{-i\nu \pi /2}\left({\rm ker}_\nu(x)-i {\rm kei}_\nu(x)\right),
\end{equation}
where ${\rm ker}_\nu$ and ${\rm kei}_\nu$ are Kelvin functions
\citep{W66}.

The second order term, $\left|\partial \bm{l} /\partial R \right|^2$, which
we choose to neglect, from equation~(\ref{int}) has a magnitude that is
largest in the disc around $R_{\rm warp}$ but is negligible for small
inclination angles. The largest error occurs when the outer disc is
inclined at an angle of $\pi/2$ to the black hole. Then the relative
magnitude of the neglected term is $0.049 \, \nu_{20}/\nu_{10}$ for
$\beta=0$. For $\beta=3$ it grows to $0.093 \, \nu_{20}/\nu_{10}$. If
$\nu_{20}<\nu_{10}$ the analysis is good for all inclinations of the outer
disc. When this inclination is reduced to $\pi/6$ the relative magnitude of
the neglected term has fallen to $0.023 \, \nu_{20}/\nu_{10}$ for $\beta=3$
and $0.012 \, \nu_{20}/\nu_{10}$ for $\beta=0$.

\section{The shape of the disc}
\label{sec:inc}

\begin{figure}
\epsfxsize=8.4cm 
\epsfbox{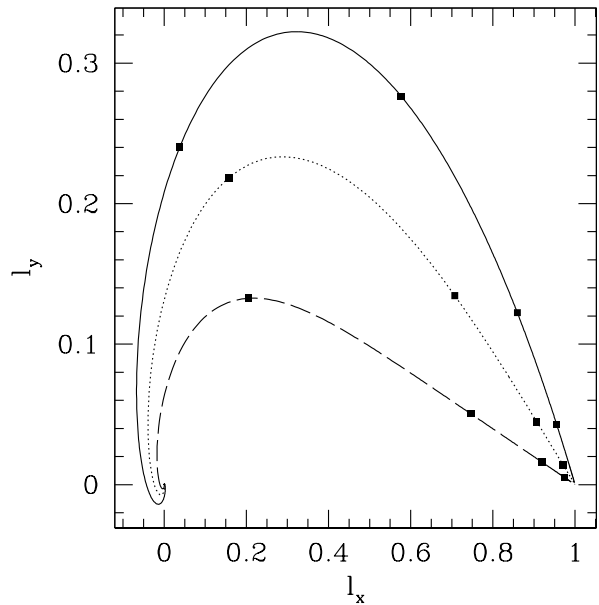}
\caption[]
{ The steady state disc, $l_y$ against $l_x$ as $R$ changes in the
  case with $W_\infty=1$. The solid line has $\beta=0$, the dotted line has
  $\beta=3/4$ and the dashed line has $\beta=3$. The lines begin at
  $R=0$ at $l_x=l_y=0$ and the dots on the curves are where $R=1$,
  $10$,$100$ and $1000 \,\rm R_{warp}$. }
\label{doub}
\end{figure}

In Figure~\ref{doub} we plot the solution $W = l_x + i l_y$ for
various values of $\beta$ as $l_x=\Re (W)$ against $l_y=\Im (W)$ with
$W_\infty=1$. Note that since the problem is a linear one, we may take
$W_\infty = 1$. In this plane, the completely flat, but inclined, disc
would be a point at $W = 1$. As $R \rightarrow 0$ the disc becomes
steadily more aligned with the black hole spin and $W \rightarrow 0$.
The lower the value of $\beta$ the less the disc is twisted. The lines
begin at $R=0$ at $l_x=l_y=0$ and the dots on the curves are where
$R/R_{\rm warp} =1$, $10$,$100$ and $1000$.  

Kelvin functions are in effect combinations of the ordinary,
oscillatory Bessel functions $J(x)$ and $Y(x)$ and the modified,
non-oscillatory Bessel functions $I(x)$ and $K(x)$. Because of the
nature of Kelvin functions and because, as $R \rightarrow 0$, the
argument of the Bessel function tends to infinity, the solution $W(R)$
circles the origin an infinite number of times as $R \rightarrow 0$
while at the same time approaching the origin exponentially. Thus, as
we approach the origin, the disc becomes very twisted, but very flat.
This explains why, as remarked by \cite{SF}, numerical integration
packages tend to fail for this problem.

The inclination of the disc relative to the black hole spin
direction $\hat{\bm{z}}$ at radius $R$ is
\begin{equation}
\theta(R)=\cos^{-1}\left(\hat{\bm{z}}.\bm{l}\right)=\cos^{-1}(l_z).
\end{equation}
The solution for $W$ is in the frame of the black hole where
$\bm{J}/J=(0,0,1)$ and we have the disc angular momentum vector
\begin{equation}
\bm{l}=\left(\Re (W), \Im (W), \sqrt{1-|W(R)|^2}\right).
\end{equation}
Thus the inclination of the disc at radius $R$ is this frame is
\begin{equation}
\theta_1(R)=\cos^{-1} \left(\sqrt{1-|W(R)|^2}\right).
\end{equation}

We can also find the disc inclination in the frame aligned with the
outer disc regions. In this frame the inclination of the disc tends to
zero as $R\rightarrow \infty$. This means as $R\rightarrow \infty$,
$\bm{l}\rightarrow (0,0,1)$. In this frame we take the angular
momentum of the black hole, $\bm{J}'$, to have $j_y'=0$ at $t=0$ and so
\begin{equation}
\bm{J'}=J\left(-\sin \eta ,0,\cos \eta \right)
\end{equation}
where $\eta$ is the angle of inclination of the black hole spin to the
outer disc axis and prime denotes quantities in the frame of the outer disc.
The disc angular momentum direction vector becomes
\begin{align}
l_x'&=l_x\cos \eta -l_z \sin \eta \cr
&= \Re (W)\cos \eta  - \sqrt{1-|W|^2}\sin \eta\cr
l_z'&=l_x\sin \eta  +l_z\cos \eta \cr
&=  \Re (W)\sin \eta + \sqrt{1-|W|^2}\cos \eta\cr
l_y'&=l_y
\end{align}
 then
\begin{equation}
\theta_2(R)=\cos^{-1}\left(\Re (W)\sin \eta  + \sqrt{1-|W(R)|^2} \cos \eta \right).
\end{equation}

If we transform back to the frame of the black hole we would need the transformation
\begin{align}
l_x&= l_x'\cos \eta +l_z'\sin \eta \cr
l_z&= -l_x'\sin \eta +l_z'\cos \eta \cr
l_y&=l_y'
\end{align}
and so we see that in the frame of the black hole as $R\rightarrow \infty$
\begin{equation}
\bm{l}\rightarrow (\sin \eta, 0, \cos \eta)
\end{equation}
because $\bm{l}'=(0,0,1)$ and so we find
\begin{equation}
W_\infty=\sin \eta,
\end{equation}
or more generally
\begin{equation}
W_\infty=-(j_x'+ij_y').
\end{equation}

\subsection{Application to GRO J1655-40}

\begin{figure}
\epsfxsize=8.4cm 
\epsfbox{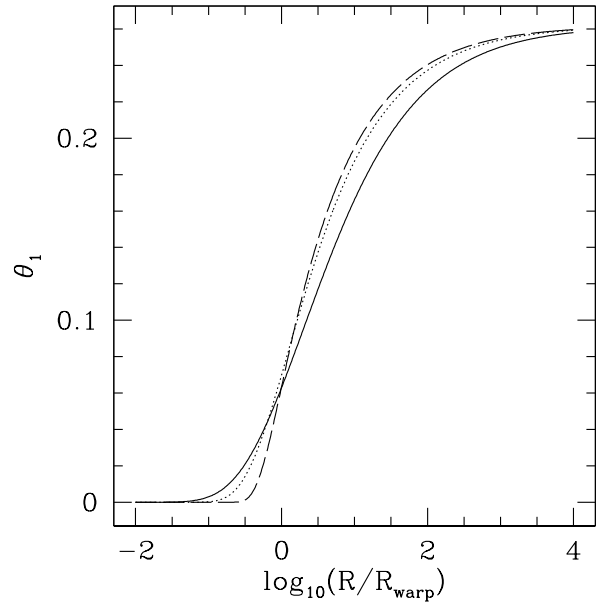}
\caption[]
{ The inclination of the disc against $R/R_{\rm warp}$ in the frame of
  the black hole. The solid line has $\beta=0$, the dotted line has
  $\beta=3/4$ and the dashed line has $\beta=3$. }
\label{inclin1}
\end{figure}

\begin{figure}
\epsfxsize=8.4cm 
\epsfbox{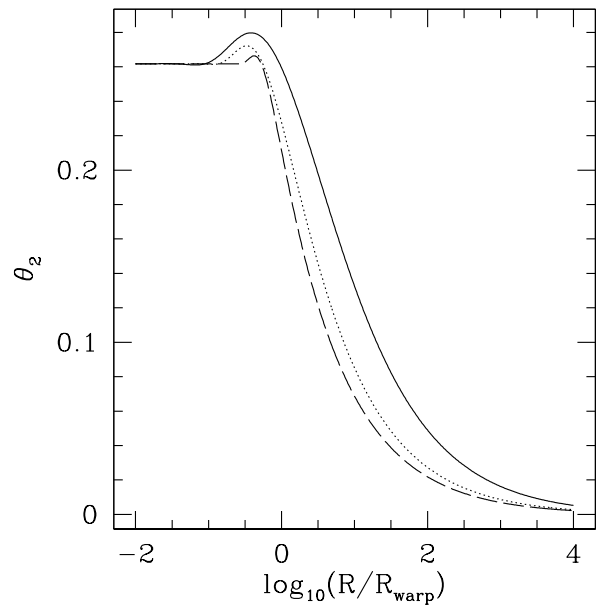}
\caption[]
{  The inclination of the disc against $R/R_{\rm warp}$ in the frame of
  the binary. The solid line has $\beta=0$, the dotted line has
  $\beta=3/4$ and the dashed line has $\beta=3$. }
\label{inclin2}
\end{figure}

As a real example we consider parameters relevant to the source GRO
J1655-40 and take $\eta = 0.2618$ corresponding to an angle of
$15^\circ$. We assume that the outer disc plane corresponds to that of
the binary and that the inner disc is aligned with the spin of the
hole which is parallel to the observed jet. In Figure~\ref{inclin1} we
plot the inclination in the frame of the black hole and in
Figure~\ref{inclin2} of the binary. The former shows that the warp
steepens with increasing $\beta$. The latter shows that, in the frame
of the binary, the inclination of the disc near $R=0$ is higher than
the inclination of the black hole. This hump, that is due to the
precession of the disc around the hole, could shield the binary
companion from the black hole radiation.

\section{Alignment and precession timescales}

Having obtained the shape of the disc we are now able to calculate the
mutual torque between it and the black hole. Because we assume the
outer disc is fixed, this enables us to compute the timescale on which
the black hole spin aligns with it and to compute the precession rate as
it does so.  We work in the frame of the black hole and the torque on
the black hole is is given by
\begin{align}
-\frac{d\bm{J}}{dt} = & \int_{\rm disc} \frac{\bm{\omega}_{\rm p} \times \bm{L}}{R^3}2\pi R\, dR \\ 
                        = & \int_{R_{\rm in}}^{R_{\rm out}} \omega_{\rm p}(-l_y,l_x,0)C_2 2 \pi R^{-\frac{3}{2}-\beta}\, dR,
\end{align}
where $R_{\rm in}$ and $R_{\rm out}$ are the inner and outer edge of
the disc and  $C_2 = (GM)^{1/2} \Sigma_0 R_{0}^\beta$ is a constant. Adding the $x$-component to $i$ times the
$y$-component we obtain
\begin{align}
\frac{d(J_x+iJ_y)}{dt}= -2\pi i \omega_{\rm p} C_2 \int_{R_{\rm in}}^{R_{\rm out}} W R^{-\frac{3}{2}-\beta}\, dR
\end{align}
and using equation~(\ref{W1}) to substitute for $W$ we find
\begin{align}
\frac{d(J_x+iJ_y)}{dt}=&-2\pi i \omega_{\rm p} C_2 BR_{0}^{\frac{1}{4}}\int_{R_{\rm in}}^{R_{\rm out}}R^{-\frac{7}{4}-\beta}\cr
&\times K_{\frac{1}{2(1+\beta)}}\left(\frac{2}{1+\beta}\kappa^{\frac{1}{2}}R^{-\frac{1}{2}(1+\beta)}\right) 
\, dR,
\end{align}
where the constant $B$ is defined by equation~(\ref{B}).  We simplify
the expression by taking dimensions out of the integral to obtain
\begin{align}
\frac{d(J_x+iJ_y)}{dt}= & \frac{4\pi i \omega_{\rm p} C_2}{1+\beta}B
R_{0}^{\frac{1}{4}}
\left( \frac{1+\beta}{2 \kappa^{\frac{1}{2}}}\right)^{\frac{3+4\beta}{2(1+\beta)}} \cr
&\times \int_{z_{\rm in}}^{z_{\rm out}} K_{\frac{1}{2(1+\beta)}}(z)z^{\frac{1+2\beta}{2(1+\beta)}}\,dz,
\label{djdt}
\end{align}
where 
\begin{equation}
z(R)=\frac{2}{1+\beta}\kappa^{\frac{1}{2}}R^{-\frac{1}{2}(1+\beta)},
\end{equation}
so that $z_{\rm in}=z(R_{\rm in})$ and $z_{\rm out}=z(R_{\rm out})$.

In Appendix~1 we show that
\begin{align}
\int_0^{(1-i)\infty} K_c(z)z^d\,dz= 2^{d-1}&\Gamma \left(\frac{1}{2}(1-c+d)\right)\cr 
&\Gamma \left(\frac{1}{2}(1+c+d)\right)
\end{align}
if $\Re(c-d)<1$ and $\Re(c+d)>-1$. So with $c=1/(2(1+\beta))$ and
$d=(1+2\beta)/(2(1+\beta))$ this integral is valid for $\beta
>-1/2$. We now have
\begin{align}
\frac{d(J_x+iJ_y)}{dt}= & \frac{4\pi i \omega_{\rm p} C_2}{1+\beta}B R_0^{\frac{1}{4}}
\left( \frac{1+\beta}{2 \kappa^{\frac{1}{2}}}\right)^{\frac{3+4\beta}{2(1+\beta)}}
                            2^{-\frac{1}{2(1+\beta)}} \cr
&\times \Gamma \left(\frac{1+2\beta}{2(1+\beta)}\right)
\end{align}
and using equation~(\ref{B}) to eliminate $B$ we get
\begin{align}
\frac{d(J_x+iJ_y)}{dt} = &\frac{2\pi i \omega_{\rm p} C_2}{1+\beta}W_\infty
\left( \frac{1+\beta}{\kappa^{\frac{1}{2}}}\right)^{\frac{1+2\beta}{1+\beta}} \cr
& \times \frac{\Gamma \left(\frac{1+2\beta}{2(1+\beta)}\right)}{\Gamma \left(\frac{1}{2(1+\beta)}\right)}.
\end{align}

In Section~\ref{sec:inc} we found that $W_\infty=-(j_x+ij_y)$. We define
\begin{equation}
\Gamma_\beta=\frac{\Gamma \left(\frac{1+2\beta}{2(1+\beta)}\right)}{\Gamma \left(\frac{1}{2(1+\beta)}\right)},
\end{equation}
so that
\begin{align}
\frac{d(j_x+ij_y)}{j_x+ij_y}=
\frac{2\pi i C_2}{1+\beta}\frac{\omega_{\rm p}}{J}
\left( \frac{1+\beta}{\kappa^{\frac{1}{2}}}\right)^{\frac{1+2\beta}{1+\beta}}
\Gamma_{x} 
 dt.
\end{align}
Using equation~(\ref{omegap}) we get
\begin{align}
\frac{d(j_x+ij_y)}{j_x+ij_y}=
\frac{4\pi i G C_2}{c^2}
(1+\beta)^{\frac{\beta}{1+\beta}}\kappa^{-\frac{1+2\beta}{2(1+\beta)}}
\Gamma_{x} dt
\end{align}
and equation~(\ref{kappa}) for $\kappa$ we get
\begin{align}
\frac{d(j_x+ij_y)}{j_x+ij_y}=
&-(-i)^{\frac{1}{2(1+\beta)}}\frac{2\pi
  (G^3M)^{1/2}\Sigma_0}{c^2}R_{0}^{\frac{\beta}{2(1+\beta)}}2^{\frac{1}{2(1+\beta)}}  \cr
&\times (1+\beta)^{\frac{\beta}{1+\beta}}\left(\frac{\omega_{\rm p}}{\nu_{20}}\right)^{-\frac{1+2\beta}{2(1+\beta)}}
\Gamma_\beta
 dt.
\end{align}
Then using equations~(\ref{omegap}) and~(\ref{angmom}) we find
\begin{align}
\frac{d(j_x+ij_y)}{j_x+ij_y}= &-(-i)^{\frac{1}{2(1+\beta)}}\frac{2\pi
  (G^3M)^{1/2}\Sigma_0}{c^2}R_{0}^{\frac{\beta}{2(1+\beta)}}
\cr
&\times \left(\frac{1+\beta}{2}\right)^{\frac{\beta}{1+\beta}}\left(\frac{aG^2M^2}{\nu_{20}
  c^3}\right)^{-\frac{1+2\beta}{2(1+\beta)}} \cr
& \times \Gamma_\beta dt.
\label{full}
\end{align}

\begin{figure}
\epsfxsize=8.4cm 
\epsfbox{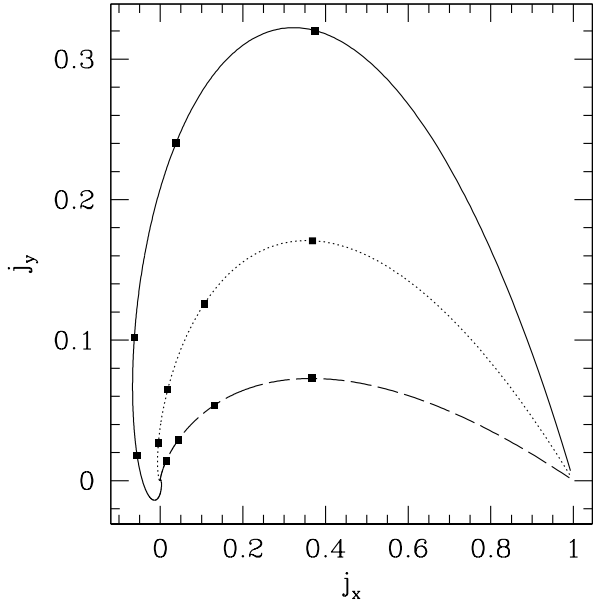}
\caption[]
{ The angular momentum of the hole, $j_y$ against $j_x$ evolving in
  time. Initially $j_x=1$ and $j_y=0$. The points along the
  lines are at times $t=1$, $2$, $3$ and~$4 \,\rm T$. }
\label{hole}
\end{figure}

We can rewrite equation~(\ref{full}) as
\begin{equation}
\frac{d(j_x+ij_y)}{j_x+ij_y}=-(-i)^{\frac{1}{2(1+\beta)}} \frac{dt}{T}
\label{time}
\end{equation}
by setting
\begin{align}
T^{-1}= &  \frac{2\pi (G^3M)^{1/2}\Sigma_0}{c^2}R_{0}^{\frac{\beta}{2(1+\beta)}} 
\left(\frac{1+\beta}{2}\right)^{\frac{\beta}{1+\beta}} \cr
&\left(\frac{aG^2M^2}{\nu_{20} c^3}\right)^{-\frac{1+2\beta}{2(1+\beta)}}
\Gamma_\beta .
\end{align}
We can then integrate to find
\begin{align}
j_x+ij_y=& A \exp \left[-(-i)^{\frac{1}{2(1+\beta)}} \frac{t}{T} \right] \cr 
 =& A\exp \left[-\cos \left(\frac{\pi}{4(1+\beta)}\right)\frac{t}{T}\right] \cr
&\times \exp \left[i\sin
    \left(\frac{\pi}{4(1+\beta)}\right)\frac{t}{T} \right],
\label{int2}
\end{align}
where $A$ is the value of $j_x+ij_y$ at $t=0$. In Figure~\ref{hole} we
plot the evolution of $j_y$ against $j_x$ with $A=1$ so $j_x=1$ and
$j_y=0$ at $t=0$. The points along the lines are at times $t=1$, $2$,
$3$ and~$4 \,\rm T$.

\subsection{Alignment Timescale}

Thus the timescale for alignment of the black hole is
\begin{align}
t_{\rm align}= & \frac{T}{\cos \left(\frac{\pi}{4(1+\beta)}\right)}\cr
=&\frac{c^2}{2\pi (G^3M)^{1/2}\Sigma_0}R_0^{-\frac{\beta}{2(1+\beta)}}
\left(\frac{2}{1+\beta}\right)^{\frac{\beta}{1+\beta}}  \cr
&\times \left(\frac{aG^2M^2}{\nu_{20} c^3}\right)^{\frac{1+2\beta}{2(1+\beta)}}
\frac{\Gamma \left(\frac{1}{2(1+\beta)}\right)}{\Gamma \left(\frac{1+2\beta}{2(1+\beta)}\right)}\cr
&\times \frac{1}{\cos \left(\frac{\pi}{4(1+\beta)}\right)}.
\end{align}
Putting $R_0=R_{\rm warp}$ we find
\begin{align}
t_{\rm align}=&\frac{c^2}{4\pi (G^3M)^{1/2}\Sigma_0}R_{\rm warp}^{1/2} 
(1+\beta )^{-\frac{\beta}{1+\beta}} \cr
&\times \frac{\Gamma \left(\frac{1}{2(1+\beta)}\right)}{\Gamma \left(\frac{1+2\beta}{2(1+\beta)}\right)\cos \left(\frac{\pi}{4(1+\beta)}\right)}.
\end{align}
If $\beta=0$ then we get
\begin{equation}
t_{\rm align}(0)= \frac{1}{\sqrt{2} \pi \Sigma_0}\left(\frac{acM}{\nu_{20}G}\right)^{\frac{1}{2}}
\end{equation}
which agrees with \cite{SF} who omitted a factor of $\sqrt{2}$.

\begin{figure}
\epsfxsize=8.4cm 
\epsfbox{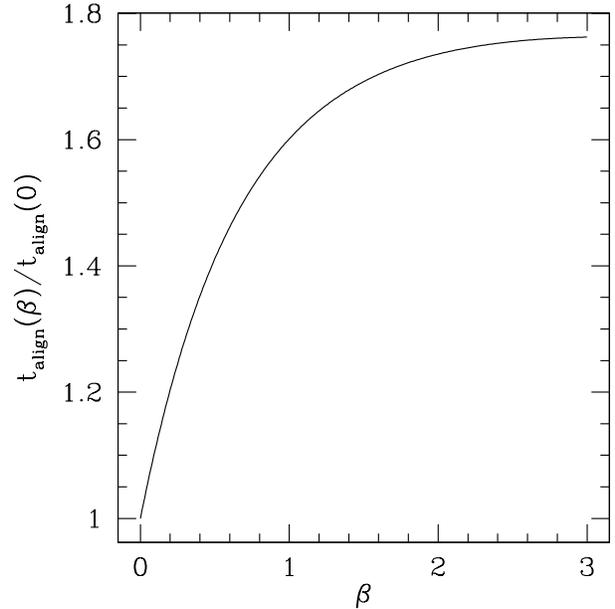}
\caption[]
{ The alignment timescale against $\beta$ normalised by the alignment
  timescale when $\beta=0$. }
\label{al}
\end{figure}

We can write the timescale to align with $\beta$ in terms of the
timescale with $\beta=0$ so that
\begin{equation}
\frac{t_{\rm align}(\beta)}{t_{\rm align}(0)}=\frac{(1+\beta)^{-\frac{\beta}{1+\beta}}}{\sqrt{2}}
\frac{\Gamma \left(\frac{1}{2(1+\beta)}\right)}{\Gamma \left(\frac{1+2\beta}{2(1+\beta)}\right)\cos \left(\frac{\pi}{4(1+\beta)}\right)}.
\label{tt0}
\end{equation}
In Figure~\ref{al} we plot this ratio as a function of $\beta$. Like
\cite{NA99}, we see that as $\beta$ increases, the timescale of
alignment increases, but only by just under a factor of 2 as $\beta$
varies from 0 to 3.

\subsection{Precession Timescale}

\begin{figure}
\epsfxsize=8.4cm 
\epsfbox{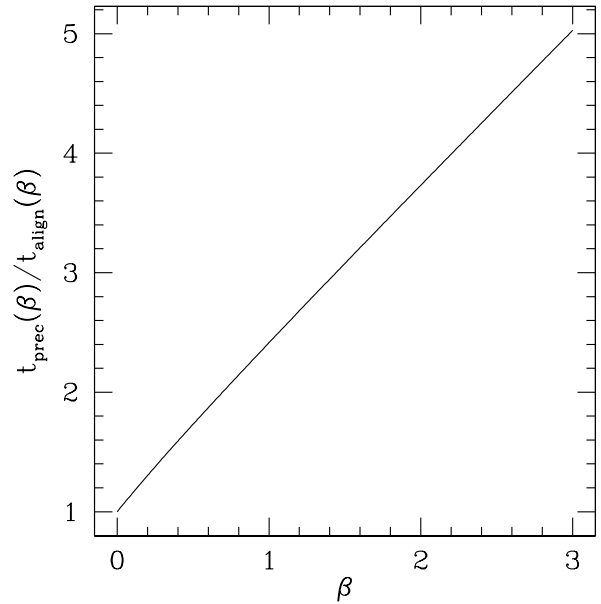}
\caption[]
{The precession timescale divided by the alignment timescale against
  $\beta$.  }
\label{prec}
\end{figure}

The precession timescale is 
\begin{align}
t_{\rm  prec}= & \frac{T}{\sin(\frac{\pi}{4(1+\beta)})} \cr
=&\frac{\cos \left(\frac{\pi}{4(1+\beta)}\right)}{\sin \left(\frac{\pi}{4(1+\beta)}\right)}t_{\rm align}\cr
=&\cot \left(\frac{\pi}{4(1+\beta)}\right)t_{\rm align}.
\end{align}
Thus, as found by \cite{SF}, when $\beta = 0 $ these two timescales
are identical.  In Figure~\ref{prec} we plot the ratio of
the precession timescale to alignment timescale against $\beta$. If
$\beta=0$ then the alignment and precession timescales are the same
but if $\beta>0$ then the precession timescale is longer than the
alignment timescale and increases with increasing $\beta$. This
is also apparent from Figure~\ref{hole}.

\section{Truncation of the Disc}
\label{sec:truncation}

The results in the preceding Section were obtained under the
assumption that the outer disc radius is infinite. We actually only
require $R_{\rm out} \gg R_{\rm warp}$. However, this may not always
be the case. For example in binary star systems the disc is truncated
at the tidal radius. We consider here the effects of truncating the
disc at finite radius.

The mass of the disc in steady state is given by
\begin{align}
M_{\rm disc}&=2\pi \int_{R_{\rm in}}^{R_{\rm out}} \Sigma R\, dR 
            =2\pi \Sigma_0 R_{0}^{\beta}\int_{R_{\rm in}}^{R_{\rm out}} R^{1-\beta}\, dR \cr
            &=\frac{2\pi \Sigma_0}{R_{0}^{-\beta}(2-\beta)}\left[ R_{\rm out}^{2-\beta}-R_{\rm in}^{2-\beta}\right]  
\end{align}
if $\beta \ne 2$. If $\beta <2$ then for finite mass $R_{\rm out}$
must be finite and if $\beta >2$ then $R_{\rm in}\ne 0$.

We consider the integral in equation~(\ref{djdt}).  If we let
$z=(1-i)y$ where $y$ is real and 
\begin{equation}
y=\frac{2}{(1+\beta)}\left(\frac{\omega_{\rm p}}{\nu_{20}}\right)^{\frac{1}{2}}R_{0}^{\frac{\beta}{2}}R^{-\frac{1}{2}(1+\beta)}
\end{equation}
then the integral becomes
\begin{align}
Q=\int_{y_{\rm in}}^{y_{\rm out}}(1-i) K_{\frac{1}{2(1+\beta)}}((1-i)y)((1-i)y)^{\frac{1+2\beta}{2(1+\beta)}}\,dy
\end{align}

We now define
\begin{align}
P=\frac{\int_{y_{\rm out}}^{\infty}(1-i) K_{\frac{1}{2(1+\beta)}}((1-i)y)((1-i)y)^{\frac{1+2\beta}{2(1+\beta)}}\,dy}
{\int_{0}^{\infty}(1-i) K_{\frac{1}{2(1+\beta)}}((1-i)y)((1-i)y)^{\frac{1+2\beta}{2(1+\beta)}}\,dy}
\end{align}
where 
\begin{equation}
y_{\rm out}=\frac{2}{1+\beta}\left(\frac{\omega_{\rm p}R_0^{\beta}}{\nu_{20}}\right)^{\frac{1}{2}}R_{\rm out}^{-\frac{1}{2}(1+\beta)}
\end{equation}
so that in units of $R_0=R_{\rm warp}$
\begin{equation}
y_{\rm out}=\frac{\sqrt{2}}{1+\beta}\left(\frac{R_{\rm out}}{R_{\rm warp}}\right)^{-\frac{1}{2}(1+\beta)}.
\end{equation}

\begin{figure}
\epsfxsize=8.4cm 
\epsfbox{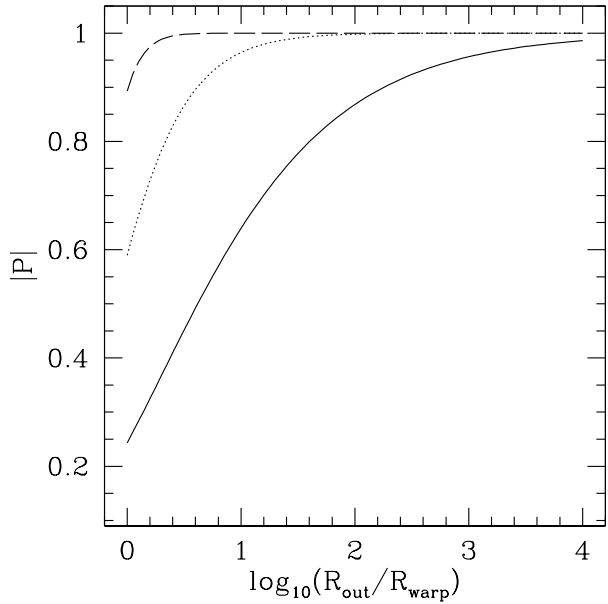}
\caption[]
{ The modulus of $P$ against $R_{\rm out}/R_{\rm warp}$. The solid
  line has $\beta=0$, the dotted line has $\beta=3/4$ and the dashed
  line has $\beta=3$.  }
\label{trunc}
\end{figure}

Thus the quantity $P(y_{\rm out})$ encapsulates the effect of
truncating the disc at radius $R_{\rm out}(y_{\rm out})$.  In
Figure~\ref{trunc} we plot the effect of truncating the disc on the
timescale for $\beta=0$,~$3/4$ and~$3$. We plot $|P|$ against $R_{\rm
  out}/R_{\rm warp}$. If $|P|=1$ then truncating the disc at that
radius has no effect on the timescale for alignment. We see that if
$\beta$ is higher, we can truncate the disc closer to the central
black hole without affecting the timescales. If $\beta=3$ we could
truncate the disc at $R_{\rm out}=4R_{\rm warp}$ whereas if $\beta=0$
we cannot truncate the disc within $R_{\rm out}\approx 10^4R_{\rm
  warp}$ and leave the timescales unchanged.

\section{Counter Alignment}

\cite{SF} considered black holes almost anti-parallel to the discs. In
this case we reverse the sign of $\omega_{\rm p}$ and obtain
\begin{equation}
\kappa =\frac{2 i \omega_{\rm p}}{\nu_{20}}R_{0}^{\beta}
\label{kappa2}
\end{equation}
and hence equation~(\ref{Wwarp}) becomes
\begin{align}
W=&\frac{2W_\infty}{\Gamma \left(\frac{1}{2(1+\beta)}\right)}\frac{(i)^{\frac{1}{4(1+\beta)}}}{ (1+\beta)^{\frac{1}{2(1+\beta)}}}\left(\frac{R_{\rm warp}}{R}\right)^{1/4} \cr
&\times K_{\frac{1}{2(1+\beta)}}\left(\frac{\sqrt{2}}{1+\beta}(1+i)\left(\frac{R}{R_{\rm warp}}\right)^{-\frac{1+\beta}{2}}\right)
\end{align}
and equation~(\ref{time}) becomes
\begin{equation}
\frac{d(j_x+ij_y)}{j_x+ij_y}=i^{\frac{1}{2(1+\beta)}} \frac{dt}{T}
\end{equation}
so that $j_x$ and $j_y$ increase exponentially and the disc realigns
initially on the same timescale.

\section{Conclusions}

We have derived the steady state profile of a warped accretion disc in
the case when the viscosity and the surface density vary as power laws
in radial distance from the central black hole.  We find that,
compared to the analysis of \cite{SF} where constant surface density
was assumed, for more realistic situations in which the surface
density is a decreasing function of radius, the timescale for
alignment of a black hole with its accretion disc increases slightly
while the timescale of precession is more greatly increased. For
constant surface density \cite {SF} found these two timescales to be
the same.  For more realistic density distributions we find that the
black hole precesses at a much slower rate than the rate at which it
aligns. If this process were responsible for changing the jet
direction in an observed source then we would predict that there
should be little evidence of precession.

\section*{Acknowledgements}

We thank Phil Armitage and Priya Natarajan for helpful comments.  RGM
thanks STFC for a Studentship. CAT thanks Churchill College for a
Fellowship.

\section*{APPENDIX 1}

\begin{figure}
\epsfxsize=8.4cm 
\epsfbox{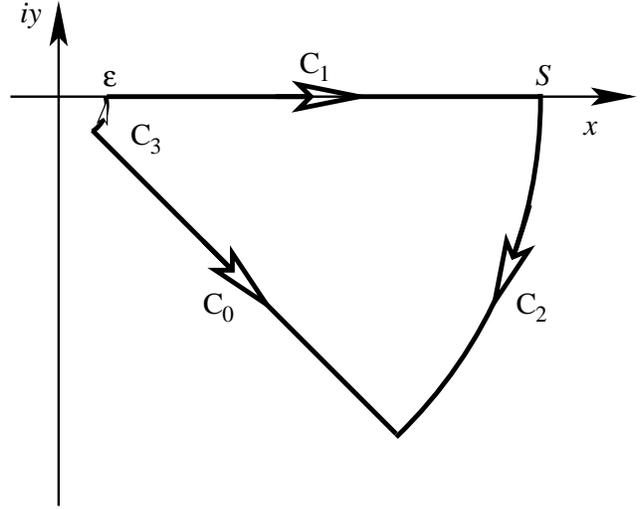}
\caption[]
{ The complex contour of integration. The contour $C_0$ is the
  integral we need. We know the integral over $C_1$. }
\label{contour}
\end{figure}

From tables of integrals \citep{GR80} we find
\begin{align}
\int_0^\infty K_c(z)z^d\,dz= 2^{d-1}&\Gamma \left(\frac{1}{2}(1-c+d)\right)\cr 
&\times \Gamma \left(\frac{1}{2}(1+c+d)\right)
\end{align}
valid for $\Re(c-d)<1$ and $\Re(c+d)>-1$.  We need the integral over
$(0,(1-i)\infty)$ along the contour $C_0$ in Figure~\ref{contour}.
There are no singularities in $K_c(z)z^d$ other than a branch point at
the origin when $d$ is not an integer.  We have
\begin{align}
Q=\int_0^{(1-i)\infty} K_c(z)z^d\,dz=\int_0^\infty K_c(z)z^d\,dz 
+Q_2 +Q_3
\end{align}
where $Q_2$ is the integral along contour $C_2$ over $z=Se^{i\theta}$
as $S\rightarrow \infty$ and $Q_3$ is the integral along contour $C_3$
over $z=\epsilon e^{i\theta}$ as $\epsilon \rightarrow 0$ with $-\pi/4
<\theta <0$. We find
\begin{align}
Q_2=\lim_{S\rightarrow \infty}\int_0^{-\pi /4} iK_c(Se^{i\theta})S^{d+1}e^{i\theta (d+1)}\,d\theta 
\end{align}
so that
\begin{align}
|Q_2| \le & \lim_{S\rightarrow \infty}\int_0^{-\pi /4} | K_c(Se^{i\theta})S^{d+1}| \, d\theta \cr
& \sim  \left( \frac{\pi}{2}\right)^{1/2}\lim_{S\rightarrow \infty} \left[S^{d+1/2}\int_0^{-\pi /4} e^{-S\cos \theta} \, d\theta \right]\cr
& \rightarrow 0
\end{align}
as $S\rightarrow \infty$ if $\cos \theta \ge 0$ so that $-\pi
/2<\theta <\pi/2$. We have used the asymptotic expansion \citep{W66}
\begin{equation}
K_c(z)\sim \left( \frac{\pi}{2z}\right)^{1/2}e^{-z}(1+...),
\end{equation}
as $z \rightarrow 0$.
We have
\begin{align}
Q_3=\lim_{\epsilon\rightarrow 0}\int^0_{-\pi /4} &iK_c(\epsilon e^{i\theta})\epsilon^{d+1}e^{i\theta (d+1)}\,d\theta \cr
\end{align}
and using the approximation in equation~(\ref{Ksmallx}) we find
\begin{align}
|Q_3| & \le \lim_{\epsilon\rightarrow 0}\left[\epsilon^{d+1}\int_{-\pi /4}^0 | K_c(\epsilon e^{i\theta})| \,d\theta \right]\cr
& \sim \frac{\pi}{4}\frac{\Gamma (c)}{2^{1-c}} \epsilon^{d-c+1}\cr
& \rightarrow 0
\end{align}
because $\Re (c-d)<1$. Thus
\begin{equation}
Q=\int_0^{(1-i)\infty} K_c(z)z^d\,dz=\int_0^\infty K_c(z)z^d\,dz .
\end{equation}

\label{lastpage}

\begin{thebibliography}{99}
  

\bibitem[\protect\citeauthoryear{Bardeen \& Petterson}{1975}]{BP}
  Bardeen J. M., Petterson J. A., 1975, ApJ, 195, L65

\bibitem[\protect\citeauthoryear{Gradshteyn \& Ryzhik}{1980}]{GR80}
Gradshteyn I. S., Ryzhik I. M., 1980, Academic Press
  
\bibitem[\protect\citeauthoryear{Greene, Bailyn \& Orosz}{2001}]{G01}
  Greene J., Bailyn C. D., Orosz J. A., 2001, ApJ, 554, 1290
  
\bibitem[\protect\citeauthoryear{Greenhill et al.}{2003}]{G03}
  Greenhill L. J., Kondratko P. T., Lovell J. E. J., Kuiper T. B.,
  Moran J. M., Jauncey D. L., Baines G. P., 2003, ApJ, 582, L11
  
\bibitem[\protect\citeauthoryear{Herrnstein, Greenhill \& Moran}{1996}]{H96} 
Herrnstein J. R., Greenhill L. J., Moran J. M.,  1996, ApJ, 468, L17

\bibitem[\protect\citeauthoryear{Hjellming \& Rupen}{1995}]{H95}
Hjellming R. M., Rupen M. P., 1995, Nat, 375, 464

\bibitem[\protect\citeauthoryear{King et al.}{2005}]{King05} 
King A. R., Lubow S. H., Ogilvie G. I., Pringle J. E., 2005, MNRAS, 363, 49

\bibitem[\protect\citeauthoryear{Kinney et al.}{2000}]{K00}
Kinney A. L., Schmitt H. R., Clarke C. J., Pringle J. E., Ulvestad J. S., Antonucci R. R. J., 2000, ApJ, 537, 152

\bibitem[\protect\citeauthoryear{Kumar \& Pringle}{1985}]{KP85}
Kumar S., Pringle J. E.,  1985, MNRAS, 213, 435

\bibitem[\protect\citeauthoryear{Lodato \& Pringle}{2006}]{LP06}
Lodato G., Pringle J. E., 2006, MNRAS, 368, 1196

\bibitem[\protect\citeauthoryear{Lodato \& Pringle}{2007}]{LP07}
Lodato G., Pringle J. E., 2007, MNRAS, submitted

\bibitem[\protect\citeauthoryear{Natarajan \& Armitage}{1999}]{NA99}
Natarajan P., Armitage P. J., 1999, MNRAS, 309, 961

\bibitem[\protect\citeauthoryear{Pringle}{1981}]{P81} 
Pringle J. E., 1981, ARA\&A, 19, 137

\bibitem[\protect\citeauthoryear{Pringle}{1992}]{P92} 
Pringle J. E., 1992, MNRAS, 258, 811

\bibitem[\protect\citeauthoryear{Scheuer \& Feiler}{1996}]{SF}
Scheuer P. A. G., Feiler R., 1996, MNRAS, 282, 291

\bibitem[\protect\citeauthoryear{Schmitt et al.}{2002}]{S02}
Schmitt H. R., Pringle J. E., Clarke C. J., Kinney A. L., 2002, ApJ, 575, 150

\bibitem[\protect\citeauthoryear{Watson}{1966}]{W66} 
Watson G. N., 1966, `A Treatise on the Theory of Bessel Functions', 2nd ed.
  Cambridge, England, CUP

\end{thebibliography}
\end{document}